# Ultrafast *In vivo* Transient Absorption Spectroscopy

Tomi K. Baikie[1*], Darius Kosmützky[2+], Joshua M. Lawrence[2,3+], Victor Gray[1,4], Christoph Schnedermann[1], Robin Horton[3], Joel D. Collins[2], Hitesh Medipally[5], Bartosz Witek[2,3], Marc M. Nowaczyk[5], Jenny Zhang[3], Laura Wey[6], Christopher J. Howe[2*], Akshay Rao[1*]

1 - Cavendish Laboratory, University of Cambridge, J. J. Thomson Avenue, Cambridge, CB3 0HE, UK
2 - Department of Biochemistry, University of Cambridge, Tennis Court Road, Cambridge, CB2 1QW, UK
3 - Yusuf Hamied Department of Chemistry, University of Cambridge Lensfield Road, Cambridge, CB2 1EW, UK
4 - Ångström Laboratory, Uppsala University, Box 523, 751 20, Uppsala, Sweden
5 - Department of Biochemistry, University of Rostock, Albert-Einstein-Str. 3, 18059 Rostock, Germany
6 - University of Turku, Department of Life Technologies, Molecular Plant Biology, Turku, 20014, Finland
+ Contributed Equally

## Abstract (150 words)

Transient absorption (TA) spectroscopy has proved fundamental to our understanding of energy and charge transfer in biological systems, allowing measurements of photoactive proteins on sub-picosecond timescales. Recently, ultrafast TA spectroscopy has been applied *in vivo*, providing sub-picosecond measurements of photosynthetic light harvesting and electron transfer processes within living photosynthetic microorganisms. The analysis of the resultant data is hindered by the number of different photoactive pigments and the associated complexity of photoactive reaction schemes within living cells. Here we show how *in vivo* ultrafast TA spectroscopy can be applied to a diverse array of organisms from the tree of life, both photosynthetic and non-photosynthetic. We have developed a series of software tools for performing global, lifetime and target analysis of *in vivo* TA datasets. These advances establish *in vivo* TA spectroscopy as a versatile technique for studying energy and charge transfer in living systems.

# Introduction

Ultrafast transient absorption (TA) has proved a vital technique in photobiological research due to its impressive sub-picosecond temporal resolution and ability to elucidate the nature of charge transport in photosynthetic systems. TA spectroscopy has revealed the ultrafast dynamics of photosynthetic light harvesting[1–3], electron transport[4–6] and photoprotection[7,8]. This research has largely been performed *in vitro*, identifying energy and charge transfer processes within the various pigment-protein complexes found across diverse phototrophic organisms, including light harvesting complexes, photosystems, rhodopsin and other photoactive biomolecules[9–16].

*In vitro* studies offer remarkable systematic clarity and naturally reduce the complexity of the photobiology by isolating specific photoactive biomolecules. However, the isolation of biomolecules from their physiological environments can lead to changes in their behaviour[17,18] and often involves time consuming extraction and purification procedures. Measuring *in vivo* ensures that the structure, interaction partners, solvation, and local dielectric properties of biomolecules are properly maintained.

However, the *in vivo* study of live cell systems with ultrafast TA spectroscopy has proved difficult due to scattering arising from cells and their extracellular polymeric substances.[19] Only recently have experimental parameters been optimised to resolve cells at room temperature, utilising small ($< 5$ μm diameter) photosynthetic microorganisms: the cyanobacterium *Synechocystis sp.* PCC 6803 (hereafter *Synechocystis*)[17] and the eukaryotic green alga *Nannochloropsis oceanica*[20,21]. TA spectroscopy has also been applied to study similarly sized thylakoids, extracted from the chloroplasts of plants [22–24]. These studies have demonstrated the validity of *in vivo* ultrafast TA spectroscopy for resolving photosynthetic processes such as photosynthetic electron transport and non-photochemical quenching at appropriate time scales, where comparison of spectra from mutants provided insights into the functions of proteins involved[17,21].

The time-resolved spectra from *in vivo* TA spectroscopy studies are complex, owing to the simultaneous measurement of a multitude of photoactive biomolecules, their overlapping spectra, and often similar lifetimes. This makes the analysis of specific processes difficult, and typically spectral features have been identified by *a priori* comparison to *in vitro* TA spectroscopy studies, or by the piecewise analysis of knockout mutants lacking specific proteins[17,24]. In general, a complete description of photoactive pathways *in vivo* is unlikely due to the sheer number and complexity of photoactive pathways being simultaneously probed. Hence model-independent analysis is useful to identify and elucidate the effects of experimental parameters.

Here, we demonstrate the widespread applicability of *in vivo* TA spectroscopy for photobiological research by performing measurements a diverse array of photosynthetic cells spread across the tree of life. For the first time we have developed both the apparatus (see **Methods 1**) and cell preparation (see **Methods 2-3, 6**) to study photoactive processes in non-phototrophic cells, as well as performing measurements of large cells (up to 20 μm in diameter). We have also developed a series of software tools for the analysis of these complex systems, from which we may understand the systems in a model-independent manner or with assumed reaction schemes.

# Results

## Performing *in vivo* TA

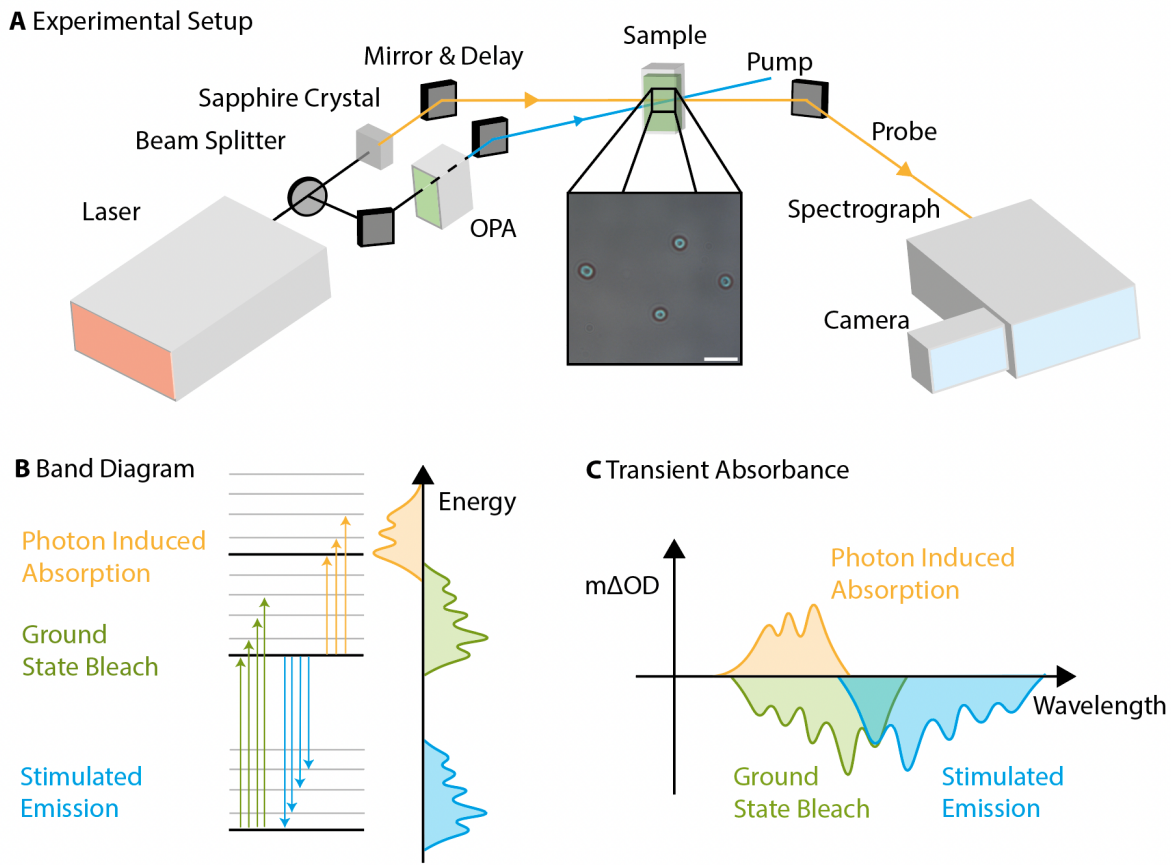

**Figure 1: In vivo transient absorbance spectroscopy**
**A -** Overview of transient absorbance (TA) setup used for in vivo TA measurements with *Synechocystis* pictured. OPA represents an optical parametric amplifier. Scale bar is 5 $\mu m$. **B -** Band diagram of theoretical energy levels transitions occurring in some biological pigment. **C -** Corresponding components of TA spectra to the band transitions in B, where the sum of the components corresponds to the resultant measured spectra.

We present an overview of our TA spectroscopy set up suitable for *in vivo* measurements. Within the cuvette containing the biological sample two pulsed beams converge; a pump which is used to excite the sample, and a broad white light supercontinuum probe beam (**Figure 1A**, for more details see **Methods 1**). The change of transmission, $\Delta T$, is obtained by comparing the transmission of the probe with and without the pump excitation. The differential transmission, $\Delta T/T$, is then recovered by dividing $\Delta T$ by $T$. A TA dataset consists of a series of difference spectra spanning several time delays of the probe relative to the pump.

As is standard in the photobiology literature, we report the data as the difference of optical density (ΔOD). ΔOD is related to differential transmission ($\Delta T/T$) by **Equation 1** (see **SI Section 1** for derivation),

$$\Delta\text{OD} = -\log_{10}\left(1 + \frac{\Delta T}{T}\right).$$

**Equation 1**

This aids interpretation of TA datasets, as ΔOD is proportional to the concentration, whereas differential transmission is not. However, care must be taken in relating differential transmission to differential absorption as these are not necessarily equivalent (see **SI Section 2** for details). Assuming data is presented in units of ΔOD, the signal may be understood in terms of the ratio in **Equation 1**. A positive signal (i.e., a positive ΔOD) suggests less transmission (i.e. a negative ΔT) and negative signal suggests more transmission at a particular wavelength[25]. The TA signal observed is generally a contribution of different signals, such as the ground state bleach (GSB), photon induced absorption (PIA) and stimulated emission (SE) from the photoactive proteins (see **Figure 1B-C** & **SI Section 1**)[13,14]. The GSB describes the situation in which pump excitation decreases the population of ground-state pigments to excited pigments in the sample, leading to a decrease in the signal. Similarly, SE describes a phenomenon in which the probe pulse induces emission of a photon from an excited pigment, leading to a decrease in the TA signal. On the other hand, PIA results in a positive signal, as upon excitation with the pump beam, optically allowed transitions from excited states to higher excited states may be accessed.

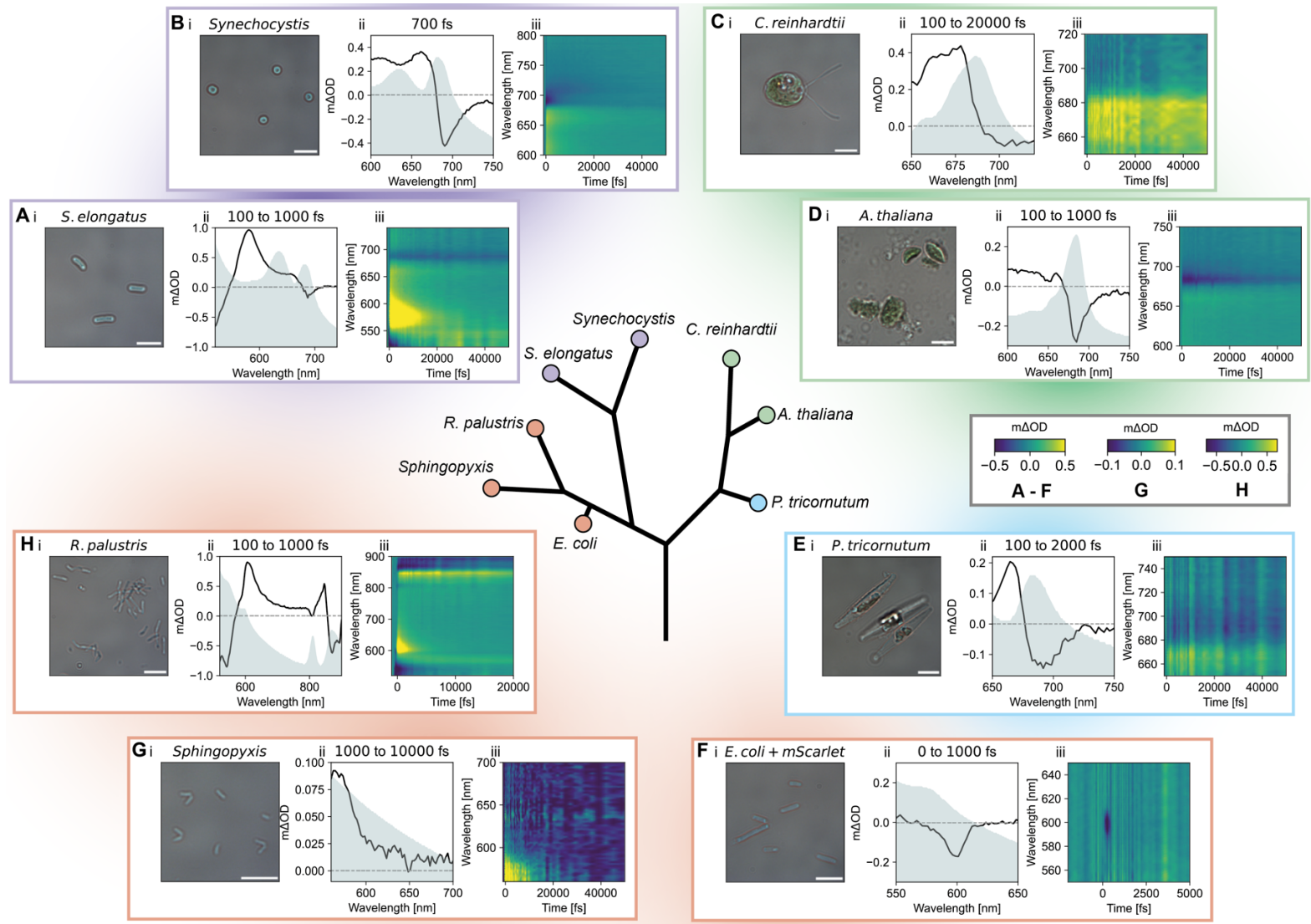

**Figure 2: Transient absorbance spectra of diverse living cells**
Sub-panels depict (i) microscopy images with 5 μm scale bar, (ii) overlayed absorbance (in solid gray, scaled for visual appeal) and TA spectra (as black line) averaged over specific time points, and (iii) full TA spectra across all wavelengths and time points. Key references colour scale to mΔOD for labelled subpanels. TA spectra were recorded of the cyanobacteria **A** - *Synechococcus elongatus* PCC 7942, **B** - *Synechocystis* sp. PCC 6803, **C** - the freshwater green alga *Chlamydomonas reinhardtii*, **D** – crude plant cell extract from *Arabidopsis thaliana*, **E** - the marine diatom *Phaeodactylum tricornutum*, **F** - genetically modified *Escherichia coli* expressing the protein fluorophore mScarlet, **G** - an isolate of *Sphingopyxis* containing carotenoids and **H -** the purple bacterium *Rhodopseudomonas palustris*.

**Biological Applicability**

To demonstrate the widespread biological applicability of the technique, we performed *in vivo* TA spectroscopy on nine different samples of cells or cell extract. These were chosen to demonstrate the effectiveness of the technique on a diverse range of pigmented organisms across the tree of life (**Figure 2**), including oxygenic and anoxygenic photosynthetic organisms as well as pigmented non-photosynthetic bacteria.

Oxygenic photosynthetic organisms studied included three species of cyanobacteria (*Synechocystis sp.* PCC 6803, *Synechococcus elongatus* PCC 7942, *Nostoc sp.* PCC 7120), a freshwater green alga (*Chlamydomonas reinhardtii*), a marine diatom (*Phaeodactylum tricornutum*), and crude plant cell extract (from *Arabidopsis thaliana*). All these samples exhibited a prominent negative feature centred at 680-710 nm corresponding to the absorbance maxima of photosystem II (PSII) and photosystem I (PSI). The decay at these wavelengths with time is associated with charge separation in the reaction centres[17]. *In vivo* TA spectroscopy of an anoxygenic photosynthetic bacterium (*Rhodopseudomonas palustris*) revealed a similar feature centred at ~ 805 nm, corresponding with the absorbance maximum of its own bacteriochlorophyll containing reaction centre (RC).

Despite these similarities, significant differences in TA spectra were observed between photosynthetic organisms. Unique spectral features are observed corresponding to species-specific photoactive biomolecules. Cyanobacterial samples (**Figure 2A,B** & **SI Figure 3**) exhibited a feature centred at 620 - 650 nm, corresponding with the absorbance maxima of their light-harvesting phycobilisome antennae. This feature is absent in the other photosynthetic samples which lack phycobilisomes (**Figure 2 C-H**) [26–29]. Even when comparing evolutionarily similar organisms, lifetime differences of TA signals can be observed. *Synechocystis* and *S. elongatus* both have a similar PSII:PSI ratio[30,31], similar hemidiscoidal phycobilisomes[31] and were cultured under identical conditions. As expected, they exhibited the same TA spectral features (with maxima and minima at ~ 620, 665 and 680 nm). However, major differences in lifetime were observed between these signals, demonstrating differences in the excitation energy and electron transfer kinetics (see **Figure 2 C-H iii**). The filamentous cyanobacterium *Nostoc* sp. PCC 7120 was also measured (**SI Figure 3**). We encountered problems probably arising from the organism's rapid sedimentation which might be overcome by flowing cells through the laser focus. The differences in lifetimes observed demonstrate the potential of *in vivo* TA spectroscopy in understanding the variability in photosynthetic processes across different species.

Beyond photosynthetic organisms, *in vivo* TA spectroscopy on *Escherichia coli,* genetically modified to express the fluorescent protein mScarlet[32] was attempted (**Figure 2F**). Fluorescent reporters are widely used in biological measurements, often in the context of *in vivo* Förster resonance energy transfer (FRET) measurements which can report the spatial proximity of proteins or protein subunits with temporal resolution in the nanosecond timescale[33]. From our measurements, the TA signal corresponded to the emission maximum of the fluorescent protein (594 nm)[32] and was observed to decay within a few hundred femtoseconds – likely corresponding to a stimulated emission process. This demonstrates that heterologous protein reporters, which have previously been studied with *in vitro* TA spectroscopy[34], can now be measured with *in vivo* TA spectroscopy.

A yellow-pigmented bacterial isolate from the genus *Sphingopyxis* (99.34% 16S RNA sequence identity to *Sphingopyxis solisilvae*)[35,36], exhibited a TA signal in the green region (~ 550 nm), congruent with the absorbance of common microbial pigments like carotenoids or terpenes. This usefully demonstrates that pigments other than chlorophylls may be investigated with *in vivo* TA spectroscopy at their endogenous concentration, even where it may be difficult to identify these absorbance maxima in the whole-cell absorbance spectra due to scattering[37] (see absorbance spectra of *Sphingopyxis* **Figure 2G**).

Our results suggest that the ultrafast TA technique is readily applicable across a large span of cell sizes and morphologies. The *Sphingopyxis* isolate is among the smallest of the investigated organisms (~ 1 µm long and 0.3 µm wide) while the diatom *P. tricornutum* (~ 20 µm long and 3 µm wide) and the alga *C. reinhardtii* (~ 9 µm long and 7 µm wide) represent the larger organisms tested here. TA spectra were successfully recorded from cells with very diverse morphologies (**Figure 2**). This confirms that spectral features are caused by photoactive biomolecules as opposed to physical processes, such as the lensing of incident light which is performed by spherical *Synechocystis* cells[38].

Experiments were performed in different media and buffer conditions (see **Methods 2,3 & 6**). Signals were also obtained from both soluble proteins such as mScarlet as well as membrane-bound pigments and proteins such as photosystems and carotenoids[39]. Despite the crude cell extraction performed on leaves of *A. thaliana* (see **Methods 2** for extraction details and **Figure 2D** for the clear heterogeneity in particle size), features of the photoactive biomolecules could still be clearly resolved.

*C. reinhardtii* (**Figure 2C**) exhibits negative phototaxis in response to high intensity light[40]. Typically, in the TA experiment we carried out 4 scans each taking 4 minutes, which were then averaged (see **Methods 1**). In the case of *C. reinhardtii*, after the 16 minutes of measurement a clear corridor was seen in the sample where cells had moved away from the laser path. This reduced the magnitude of the signal throughout the scan sequence, and well as the signal-to-noise ratio, but did not alter the dynamics (see **SI Section 3**), so we reported the averaged scan in **Figure 2**. Similar issues associated with burning or degradation of solid-state films or pigments self-assembled on surfaces are relatively common in other experimental systems and are typically overcome by either continuously moving or flowing the sample throughout the measurement.

To illustrate the associated insights from *in vitro* and *in vivo* analysis, **Figure 3 A-D** gives the associated TA data from *Synechocystis* and the isolated reaction centres (RC) PSII and PSI. Whilst the isolated photosystems investigated here stem from a different cyanobacterium (*Thermosynechococcus vestitus BP-1*), the high conservation of photosystem I and II reaction centres across photosynthetic organisms[41] allows comparison of their energy transfer dynamics[42]. For example, previous transient absorption studies comparing PSII from plants and cyanobacteria showed remarkably similar spectra features and lifetimes[43].

Moving from *in vitro* to *in vivo* TA spectroscopy (see **Figure 3**) seems to result in increased complexity, but with the benefit of dynamics that maybe more closely resemble the impressive natural energy management of the cell[44], without altering the local environment, for example through use of detergents in *in vitro* studies[45]. An obvious similarity between the *in vitro* and *in vivo* samples is observed in the spectrum of PSI and *Synechocystis* cells between 5 - 10 ps and 10 – 50 ps. Here, the whole cell PSI signal closely matches that of PSI, as does the longer wavelengths between 700 – 715 nm.

However, the same similarities are not observed in the PSII signal where the long-lived PSII signal is not obvious in the *Synechocystis* spectrum, maps or kinetics. Through analysis of the *in vivo* dataset the PSII contribution is seen to be present and perhaps best observed in the singular value decomposition (SVD) decomposition (see **SI Figure 3**). Nevertheless, PSII exhibits much faster kinetics than the *in vitro* sample, which is likely associated with the availability of natural acceptors in the photosynthetic electron transport chain. We conclude that comparison with *in vitro* analysis will remain useful in understanding *in vivo* TA spectroscopy, but some phenomena may only be observed *in vivo.*

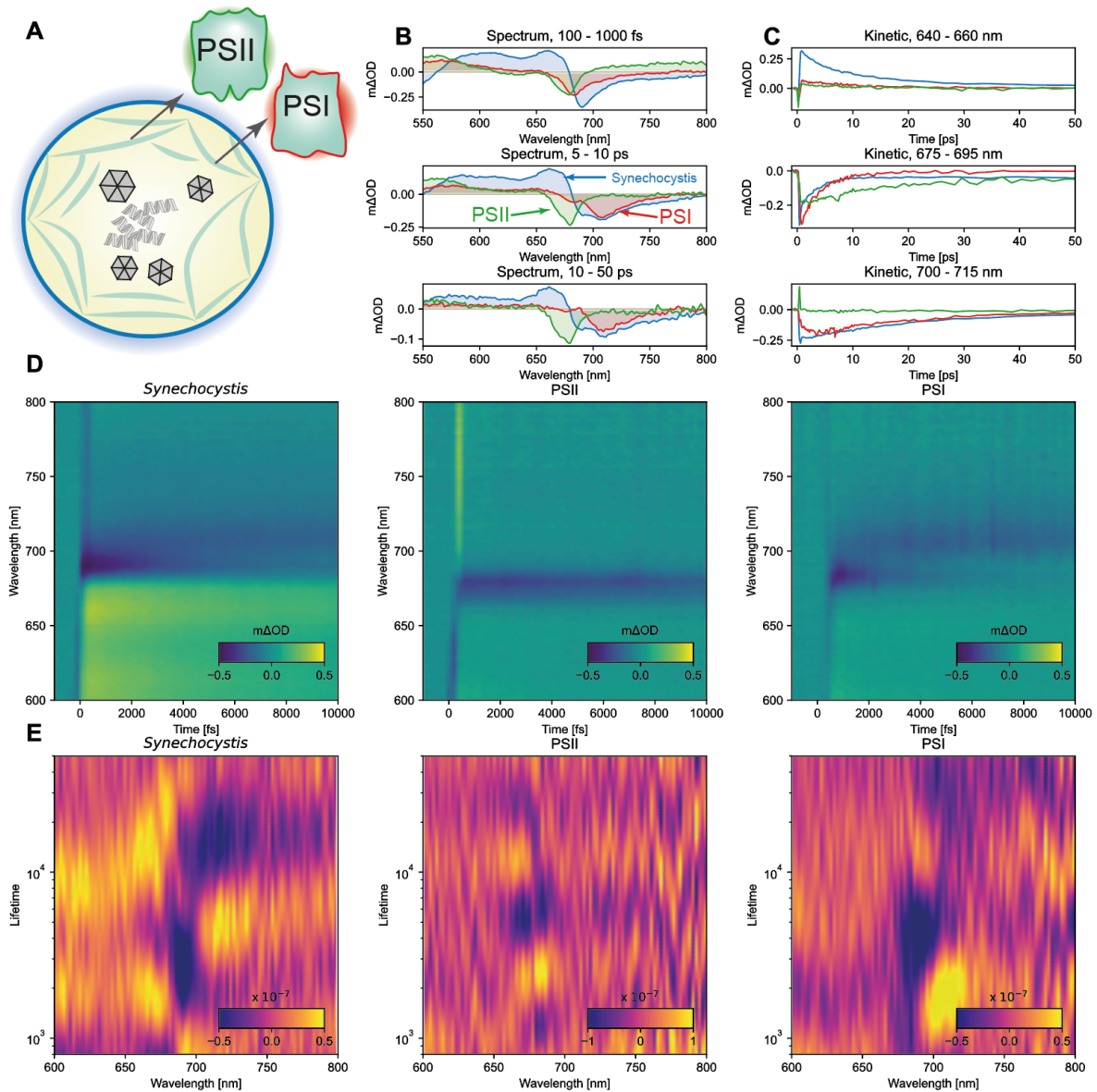

*Figure 3: In-vivo and in-vitro measurements. A* - Summary of experimental datasets from Synechocystis sp. PCC 6803 cells, isolated photosystem II from Thermosynechococcus vestitus BP-1, and isolated photosystem I from Thermosynechococcus vestitus BP-1. ***B&C –*** Spectrums and kinetics of TA results from Synechocystis (in blue), PSII (in green) and PSI (in blue). ***D*** – TA maps and ***E*** - Lifetime density maps of TA datasets. Negative values (purple) represent signal decay of a state at that lifetime whereas positive values (orange) represent signal growth of a state at that lifetime. Data is representative of n=5 biological replicates.

## Analysis Software

In understanding the photoactive system from the measured TA dataset, simultaneous analysis (meaning an analysis of all wavelengths at all times) is essential. To facilitate the analysis of these processes we have developed an open-source software package (BioTA Data Analysis – BioTADA) allowing for the simultaneous analysis of TA datasets, particularly those from *in vivo* TA spectroscopy.

BioTADA allows simultaneous analysis to be performed through a series of distinct techniques: single value decomposition (SVD), which is a convenient way to recast the transient data into temporal and spectral components, at the expense of direct physical interpretation[46]; global analysis (GA), which assumes a set number of independently decaying exponential functions; lifetime density analysis (LDA), which assumes a semi-continuous distribution of lifetimes from which lifetime distributions may be resolved; and compartment analysis (CA), which fits data to some chosen kinetic model. Comprehensive reviews of these analysis techniques have been published elsewhere[47,48].

Whilst GA and LDA require no *a priori* knowledge of a reaction scheme they cannot give physically meaningful lifetimes of excited states like CA can, unless the system is well represented by a series of independent, non-interacting components. For this reason, CA has been widely applied to *in vitro* biological samples, such as isolated photosynthetic reaction centres[15]. However, often due to the general complexity of *in vivo* TA spectroscopy, it is much harder to infer an appropriate kinetic model in samples of whole cells. Our software allows for a combined approach, where multiple analysis algorithms can be easily applied to the same dataset.

**Example Datasets**

In order to compare the appropriateness of each algorithm to analyse biological TA datasets, here we have applied the software package to the analysis of a series of example datasets. We utilise synthetic data generated using reaction schemes which are designed as illustrative examples (see **Methods 8** for details) to highlight the advantages and pitfalls of each analysis technique.

The first of these synthetic datasets introduces a simple sequential decay with three populations where exciting into one species leads to the generation of another (**Methods Figure 1A**, **Methods Equation 5**), which has been reported as representative of some photosynthetic excitation energy and electron transfer pathways[49–53] as well as FRET performed by fluorescent proteins[54]. The second dataset introduces a single moving decay with an increasing red-shift over time (**Methods Figure 1B**, **Methods Equation 6**), as has been observed in the fast red-shifts of archaeal rhodopsins[55], photosystem I[56] and chlorophyll *f*-containing photosystem II[49]. The final dataset introduces a branching decay where a population diverges into two sub-populations (**Methods Figure 1C**, **Methods Equation 7**), which is observed in photosynthetic antennae complexes, such as phycobilisomes in cyanobacteria which can perform excitation energy transfer to different reaction centres[29], or when FRET sensors are applied to unravel multi-protein interactions[57]. All the artificial datasets have had artificial noise included. Further, we performed some analysis on the *in vivo Synechocystis* (n=5) data and *in vitro* isolated photosynthetic reaction centre data discussed previously (**Figure 3D**) (n=5).

**Separability & Single Value Decomposition**

The ability to separate TA datasets into individual spectral components, representative of photoactive components, aids its interpretability. Supposing each of these spectral components behaves independently, and the measured response is linear with intensity, Beer-Lambert's law states that a noise-free, time-resolved absorption spectrum is a superposition of each of the component's absorption spectra weighted by their concentration. Such a system is therefore termed separable[46] (see **SI Section 5** for more details).

Assuming separability, TA datasets may therefore be separated into time-independent absorbance and time-dependent concentration spectra. The separability of TA datasets into time varying concentration and time independent spectra is not always valid, for example if vibrational cooling leads to a time-dependent change of a spectral feature, in which case this would need to be modelled explicitly and is demonstrated here through the moving decay synthetic dataset. To what extent a system is separable will influence the usefulness of the decomposition of the TA dataset using singular value decomposition (SVD).

Usefully, using SVD it is possible to reduce the dimensionality of the TA dataset by fitting to a noise-reduced subset, meaning instead of considering every timepoint and wavelength independently, we may choose a number of basis vectors to describe the entirety of the dataset. SVD analysis was performed on the simple linear decay, which broadly correspond to the 3 component spectra (see **SI Figure 4A**). However, the SVD analysis on the moving spectra (**SI Figure 4B**), highlights why SVD is not particularly informative in a system with a time-dependent spectral shift. As we have assumed time-wavelength separability, when the response of a single photoactive molecule shows a spectral shift, the estimated SVD components will find a number of components that exceeds the true number of compounds and thus confuse the interpretation of SVD (see **SI Section 6** for more detailed analysis).

**Global Analysis**

In global analysis (GA), we again conduct a simultaneous analysis of the dataset, or a noise-reduced subset of the dataset generated with SVD. GA is a minimally descriptive model which supposes that each spectral component decays exponentially and independently of one another, an admittedly unlikely case in the interplay of the photosynthetic systems, but nevertheless useful in obtaining lifetimes in the absence of a more sophisticated model. Further, GA provides no constraint on negative populations, which is clearly unphysical. For details on the implementation applied here see **Methods 9**. However, an important distinction of GA from SVD is that while SVD is a mathematical decomposition with arbitrary accuracy (although no canonical method for constructing such a basis exists – see **Methods 7**), the GA implementation here utilises simulated annealing targeting global minimization[58]. This means for an arbitrary dataset, there is no guarantee of finding a global minimum in finite time. Although the minimally descriptive model lessens the risk of over parameterisation, it is likely that for many biological TA datasets GA leads to an oversimplification of the system where we resolve only the minimum number of physical processes required to describe the dataset.

**Lifetime Density Analysis**

Unlike GA, lifetime density analysis (LDA) can describe systems with dynamic spectral shifts or lifetime distributions, which may be expected of complex *in vivo* systems[59].

Already, LDA has proved to be an important tool in evaluating photosynthetic energy transfer in isolated proteins[50,60–63]. Whereas GA approximates the data to the sum of a small number of exponential decays, typically on the order of 2-4, LDA in our implementation uses a large logarithmically distributed set of lifetimes[59]. Whilst LDA can be used to resolve multiple lifetimes distributed over many orders of magnitude, over-fitting of the data may complicate the interpretation (see **Methods 10**).

LDA was performed on the TA spectra of the synthetic datasets and the example photosynthetic samples (**Figure 3E** & **Methods Figure 2**), with the resultant lifetime density maps of the synthetic datasets providing accurate lifetimes for each of the individual compartments which were used to generate these datasets. This is perhaps best visualised for the moving decay dataset, where a banded formation shows the sequential formation and decay of species with increasingly red-shifted wavelengths (**Methods Figure 2B**). This demonstrates that LDA can be used to understand biological processes described by different kinetic models.

In the case of the photosynthetic samples (**Figure 3E**), lifetime density maps reveal broad spots, suggesting components have a range of underlying lifetimes. Components are observed that show rapid signal rises on the short wavelength side followed by signal decay on the long wavelength side (in the *Synechocystis* map either side of ~ 710 nm), as is typical for energy transfer from higher to lower energy pigments. The spectral features and their associated lifetimes of *Synechocystis* cells closely resembled those from PSII (see 670-680 nm) and PSI (see 680-705 nm region), which has previously been observed[17]. The lifetime density map of *Synechocystis* also shows numerous unique spectral features such as the short-lifetime features in 600-650 nm region (**Figure 3E)**, which likely correspond to the phycobilisomes[28]. LDA is an effective tool for analysing complex *in vivo* TA spectra over many complex systems, even if no model is known *a priori*.

**Compartment Analysis**

Whilst GA and LDA only model independent parallel exponential decays, compartment analysis (CA) can fit any arbitrary kinetic model, which allows for a physically relevant description of the system being studied. Typically, the process to identify a model remains an iterative approach requiring *a priori* knowledge of the system. We allow the user to define any arbitrary model, based on any number of differential equations, as long as the equations may be solved numerically (**Methods 11**). Further, any constraint, such as having only positive populations, may be applied here. This implementation has specific advantages over kinetic matrix methods which have previously been used to build kinetic models by allowing for the description of second order population dependence, which is essential for modelling excitonic processes[64] such as those which arise in photoprotection mechanisms.

The ability of CA to analyse complex TA spectra was determined by applying it to the synthetic datasets, utilising the same kinetic models which were used to generate these datasets. For the sequential decay datasets, CA analysis provided a good fit, recovering species associated spectra (SAS) which closely matches the component spectra used to generate the dataset (**Methods Figure 3**). Furthermore, the returned lifetimes are close to that of the generated scheme (see **SI Table 1** for details). However, when CA was performed again on the sequential decay dataset, this time with an incorrect kinetic model (see **Methods Figure 3D**), good fits were still obtained even though the physical interpretation of the data is clearly incorrect (see **SI Table 1**). This highlights the limitations of CA; whilst CA can be used to analyse TA datasets accurately it is important to ensure the kinetic model is justified as the use of many free parameters will likely allow a good fit to many models, with no correlation as to how well it reflects the physical reality of the system being studied[65,66]. An elephant may be drawn with only four free parameters[66], highlighting the challenge faced when exploring large parameter spaces which may result in poor quality physical insights, even with good quality fits.

Further, to complicate matters, a structurally identifiable model may be numerically unidentifiable, meaning that although the model may be correct it may be impossible to identify each compartment. In the example of the branched model, as $k_3$ and $k_4$ are similar in magnitude, we observe that the compartment analysis is unable to differentiate the two branches, as highlighted in **Methods Figure 3C**. This arises due to the similar magnitude of $k_3$ and $k_4$, meaning CA is unable to differentiate between the branching compartments.

If the spectra are fully known *a priori*, our method is readily extensible by imposing a specific spectrum as a compartment. This is particularly useful where either the spectra are entirely known, or where the separability of the concentrations and spectrum no longer is valid, as in the moving spectrum. In order to demonstrate this approach using our software package, we performed CA on the moving decay dataset by defining evolution-associated spectra (EAS)[67] consisting of a moving, decaying gaussian spectrum. This results in a good quality fit to the moving decay spectrum, recovering comparable SAS and lifetime (**Methods Figure 3B**).

Once an appropriate kinetic model has been chosen, it must be tested for statistical validity. Initially, visual inspection of residual traces or residual maps are helpful to reject deficient models, as in **SI Figure 5C**. As the noise variance is generally unknown $\chi^2$ statistics are unhelpful[46]. However, in the assumption of the presence of random noise, any structure visible on the residual map reveals a problem in the model and suggests the need for possible corrections.

**Discussion (400 words max)**

We envisage that ultrafast *in vivo* TA spectroscopy will become an essential technique for understanding ultrafast biological processes. We have demonstrated that the technique can be applied to a wide array of pigmented organisms from the tree of life. Our experimental methodology has been optimised for the study of photosynthetic organisms, which, whilst varying in size, all contain a high cellular concentration of photoactive components. However, we have also demonstrated that *in vivo* TA spectroscopy can be successfully applied to fluorescent proteins and non-photosynthetic bacteria. *In vivo* TA spectroscopy can similarly be applied to understand ultrafast charge transfer, where living cells are paired with artificial electron acceptors,[17,34] dyes[68] and photosensitisers[69].

As discussed here, global analysis techniques are ubiquitous tools for TA spectroscopists. However, characterising uncertainty in the kinetic parameters and component spectra derived from these fitting procedures is challenging. Recently, Markov chain Monte Carlo samplers have been introduced to visualise and understand uncertainty in the model fits[70]. Machine learning methods have also been used to determine the decay rate distribution of an arbitrary system without any *a priori* assumptions[71,72]. In both the machine learning and Monte Carlo methods, the strength of modelling without first establishing a model is itself an inherent problem as the physical or biophysical origins of any such model are not immediately obvious.

Simultaneous analysis is essential to resolve the details and model physical behaviour of populations in ultrafast spectroscopy experiments. The analysis methods presented here can help to significantly reduce all time dependent data to time-independent spectra and their associated time constants and facilitate its interpretation. Here we present a software toolbox that permits the modelling of any time-dependent data through established and new approaches. Throughout this work, imagined and real-world examples are used to illustrate experiments and common pitfalls. All code has been provided open source in an easy-to-use format. We believe that *in vivo* TA spectroscopy opens exciting avenues and on the horizon are studies observing elementary processes, chemical reactions, charge transfer and more, at the cellular level.

# Methods

1. **Transient Absorption Experimental Details**

In our set-up, a ytterbium doped potassium gadolinium tungstate (Yb:KGW) laser (Pharos, Light conversion) generates ≈ 200 fs narrow band pulses at 1030 nm with a repetition rate of 38 kHz. Other implementations may be used, for example Titanium Sapphire lasers, as long as they offer a sufficiently stable output.

The fundamental beam is split into two separate paths, one for the probe and the other for the pump. The pump pulse enters an optical parametric amplifier (OPA) (Orpheus, Light Conversion) from which we can select an excitation wavelength. The probe pulse enters a 3 mm sapphire crystal (Thorlabs), from which a white light supercontinuum is produced[73]. The white light probe is then mechanically delayed (Newport) with respect to the pump and a chopper (Thorlabs MC2000) which blocks a portion of the pump pulses. Both the pump and probe are focused on the sample, ensuring that the probe pulse (90 $\mu$m FWHM) is smaller than the pump pulse (250 $\mu$m FWHM) so that the area probed is uniformly excited.

The temporal resolution of the measurement is defined by the duration of the pump pulse and is ~ 200 fs in our setup. Utilising a spectrograph (Semrock 163, Oxford Instruments) equipped with a CMOS line-camera (JAI, SW2000M) we record the difference between the white light spectrum with and without pump pulses incident on the sample in a single-shot fashion at specific delays after pump illumination.

Throughout this paper we have assumed that the input data has been corrected for the instrument response function (IRF) and chirp, which arises from the frequency-dependent refractive index in the optical components. The specific nature of the chirp and IRF will be dependent on the specific experimental setup and both effects are easy to correct for[14,47,59].

2. **Culture Growth Conditions**

*Synechocystis* sp. PCC 6803, *Synechococcus elongatus* PCC 7942 and *Nostoc* sp. PCC 7120 were grown to an $OD_{750}$ of 1.0 in a shaking incubator (30 °C, 120 rpm, 40 $\mu E$ $m^{-2}$ $s^{-1}$ of continuous white light) in BG-11 medium (pH 7.8) supplemented with 10 mM sodium bicarbonate[74]. *Rhodopseudomonas palustris* CGA009 was grown to an $OD_{660}$ of 0.6 in a shaking incubator (30 °C, 120 rpm, 20 $\mu E$ $m^{-2}$ $s^{-1}$ of continuous warm white light) in Van Niel's yeast medium supplemented with 50 mM glycerol[75]. Genetically modified *Escherichia coli* DH5α was grown to an $OD_{600}$ of 1.4 in a shaking incubator (37 °C, 120 rpm) in LB medium supplemented with 100 $\mu g$ $mL^{-1}$ of carbenicillin. The Sphingopyxis isolate (99.34% 16S RNA sequence identity with *Sphingopyxis solisilvae* strain R366 (NR_157002.1), see **SI Figure 2**) was obtained as a contaminating colony growing alongside *Synechocystis* on BG-11 agar plates supplemented with 5 % LB medium. The *Sphingopyxis* isolate was grown to an $OD_{600}$ of 1.5 in a shaking incubator (30 °C, 120 rpm) in LB media. *Chlamydomonas reinhardtii* cc-124 was grown to an $OD_{750}$ of 1.5 in a shaking incubator (30 °C, 120 rpm, 40 $\mu E$ $m^{-2}$ $s^{-1}$ of continuous white light) in TAP media. *Phaeodactylum tricornutum* CCAP 1055/1 was grown to an OD of 0.8 in a shaking incubator (18ºC, 110 rpm, 50 $\mu E$ $m^{-2}$ $s^{-1}$ in a 16 h light/8 h dark cycle) in f/2 medium. *Arabidopsis thaliana* col-8 was grown in a plant growth chamber at 22 °C in 16 h light/8 h dark conditions for 5 weeks.

### 3. Preparation of mutant *E. coli*

MoClo cloning suites were used to generate an mScarlet expression plasmid with the following parts: acceptor, pICH47732, level 1 acceptor position 1[76]; Prom + 5U, pC0.034, J23104[77]; CDS, mScarlet[32]; 3U + Ter, pC0.082, TrrnB[77]. Chemically competent *E. coli* DH5α were prepared by the Inoue method[78]. Golden gate plasmid assembly and transformation was performed as described previously[79]. Following transformation, successful mutants were selected for by their characteristic pink colour and fluorescence under a transilluminator.

### 4. Analysis of Sphingopyxis isolate

Pigment analysis of the Sphingopyxis isolate was performed by taking 5 mL of culture at an $OD_{600}$ of 1.5 culture, pelleting by centrifugation (x3000 g, 10 mins), and resuspending in 600 µL of methanol. Sample was then sonicated for 1 hour at 4 °C, after which cell debris was pelleted by centrifugation (x15000 g, 10 mins). An absorbance spectrum was then collected.

Sanger sequencing of the 16S rRNA gene was performed with a colony PCR product amplified with the primers 27F (5'-AGAGTTTGATCCTGGCTCAG-3') and 1492R (5'-GGTCGCCTTGGTAGGCTT-3').

### 5. Microscopy

7 µL of cultures or cell extracts were deposited on a microscope slide, covered with a coverslip and then imaged with a light microscope (Olympus BX61) at 1000-fold magnification using a 100 X oil immersion lens.

## 6. Biological Sample Preparation for *in vivo* TA

Prior to experiments, absorbance spectra of all samples between 375-1000 nm were recorded with a UV-1800 Spectrophotometer (Shimadzu). Samples were then adjusted to achieve a maximum absorbance in this range of ~ 2, either by dilution in fresh media, or by concentration via centrifugation at 3000x g for 10 mins followed by removal of excess media. The *Arabidopsis thaliana* sample was prepared by crushing leaves with pestle and mortar in ice-cold sucrose medium (0.38 M sucrose, 0.07 M $K_2HPO_4$, 0.01 M KCl adjusted to pH 7.5) and filtering through a layer of Miracloth. The sample was subsequently kept cold and dark until the TA measurement.

## 7. Singular Value Decomposition

Assuming that transient data is recorded as a $w \times t$ matrix, $D$, where $w$ is the number of wavelengths in each spectrum and $t$ is the number of time delays at which spectra are obtained. Beer's law (**SI Equation 4**) can then be simply expressed in matrix form as the product of two matrices,

$$D = AC,$$

*Methods Equation 1*

where $A$ is a $w \times n$ matrix containing the spectra of the $n$ individual components in its columns, while $C$ is an $n \times t$ matrix containing the time behavior of the excited-state concentrations of the components in its rows.

SVD was applied to TA datasets represented by matrix $D$ (**Methods Equation 1**). If $D$ is an $m \times n$ real matrix, then $D$ can be written in the form,

$$D = USV^T.$$

*Methods Equation 2*

Where $U$ has dimensions of $m \times m$, $S$ has $m \times n$, and $V$ has $n \times n$. $S$ has entries only along the diagonal, known as singular values. The weighted left singular values (wLSV) are given by $US$ and the corresponding weighted right singular values (wRSV) by $SV^T$ in which $U$ gives the temporal dependence of the signal and $V^T$ gives the spectral dependence of the signal. The goal is to determine what subset of the data is required to adequately describe the full dataset.

Retaining the number of wLSVs whose corresponding singular values are significantly larger than the rest is a good approximation for a sufficient number of bases to retain[59]. It is also possible to avoid fitting to noise by selecting a basis which reduces permutations from noise[47]. The wLSVs can be visually inspected by the user before fitting. Noise components can have singular values larger than small signal components, so the basis chosen need not be sequential. Components corresponding to signal components should have a regular structure. The number of basis vectors chosen gives only an estimate of the number of orthogonal components appropriate to represent the data and makes no estimation of the physical or biological nature of any underlying reaction scheme.

Direct comparison of matrices $V$ and $U$ between samples are not trivial as there exists no canonical method for constructing such a basis. However, the convention, as it is implemented here, is to list the singular values, $S$, in descending magnitude. Thus, the diagonal matrix $S$ is uniquely determined by $D$, although $U$ and $V$ are not.

## 8. Synthetic Datasets

The sequential decay dataset was generated with **Methods Equation 5**, without loss of generality, we assume each spectral component is positive.

$$\frac{da(t)}{dt} = -k_1 a(t)$$

$$\frac{db(t)}{dt} = k_1 a(t) - k_3 b(t)$$

$$\frac{dc(t)}{dt} = k_2 a(t) - k_4 c(t)$$

*Methods Equation 5*

where $a(\lambda)$ is a normal distribution centred at 680 nm, $b(\lambda)$ is a combination of two normal distributions centred at 650 and 710 nm, and $c(\lambda)$ is a combination of three normal distributions centred at 770, 785 and 792 nm. Random noise is artificially generated by adding noise to each element in matrix $\mathrm{D}$. Decay constants ($k_{1,2,3,4}$) are given in **SI Table 1**.

The moving decay dataset was generated following **Methods Equation 6**,

$$\frac{da(t)}{dt} = -k_1 a(t) - k_2 a(t) - k_3 a(t),$$

*Methods Equation 6*

where the spectral component, $a(\lambda, t)$ is given by a normal distribution centred at 650 nm at $t = 0$, whereupon both the centre and standard deviation are a function of $t$, red shifting and spreading with time. Moving spectra, such as this one, are a common feature of biological samples, such as the early red shift observed in PSI.

The branching decay dataset was generated using **Methods Equation 7**

$$\frac{da(t)}{dt} = -k_1\, a(t)$$

$$\frac{db(t)}{dt} = k_1\, a(t) - 2\, k_2\, b(t)$$

$$\frac{dc(t)}{dt} = k_2\, b(t) - k_3\, c(t)$$

$$\frac{dd(t)}{dt} = k_2\, b(t) - k_4\, d(t)$$

<div align="right">*Methods Equation 7*</div>

where $a(\lambda)$, $b(\lambda)$ and $c(\lambda)$ are defined as in **Methods Equation 5**, while $d(\lambda)$ is a narrow normal distribution centred at 625 nm.

To ensure robustness of analysis algorithms, each synthetic dataset was modified with random noise which was artificially generated and added to each element in matrix D.

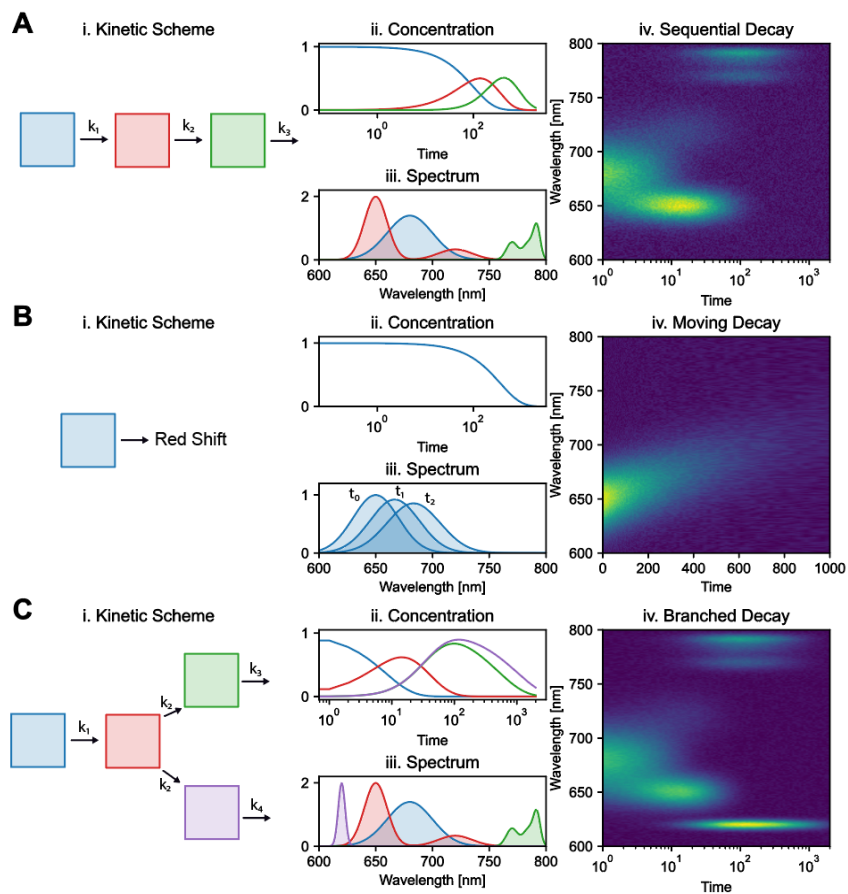

**Methods Figure 1** Synthetic datasets generated using **A-** a sequential decay model, **B-** a moving decay model, and **C-** a branching decay model. Sub-panels depict (i) kinetic models, (ii) compartment concentrations over time, (iii) species associated spectra, and (iv) transient absorbance spectra.

## 9. Global Analysis

GA was performed assuming a model of the sum of weighted exponentials,

$$\psi(t, \lambda) = \sum_i A_i(\lambda) e^{-t/\tau_i},$$

*Methods Equation 8*

where $i$ corresponds to the size of the basis expansion used to describe the data, $A_i(\lambda)$ denotes the coefficient related to each wavelength and $\tau_i$ is the lifetime associated with each exponential decay.

Following the treatment of Dorlhiac *et al.*, **Methods Equation 8** was adapted into matrix form using the reduced weighted left singular vectors in place of the full data matrix,

$$(US)_n = E(\vec{\tau})x,$$

*Methods Equation 9*

where the $(US)_n$ represents the matrix of $n$ weighted left singular vectors, $E$ is the design matrix (an exponential function dependent on the vector of lifetimes, $\vec{\tau}$), and array $x$ corresponds to $A_i$ (the coefficient for each weighted left singular vector).

A simulated annealing algorithm targeting global minimisation was used to determine which vector of lifetimes ($\vec{\tau}$) best satisfies **Methods Equation 9**[47,58]. $\vec{\tau}$ values determined from the optimization algorithm were iteratively fed to **Methods Equation 9** which was solved for $x$ by solving the associated least squares problem,

$$\min_{\vec{\tau}} |(US)_n - E(\vec{\tau})x|_2^2,$$

*Methods Equation 10*

where the subscript refers to the Euclidean norm.

# 10. Lifetime Density Analysis

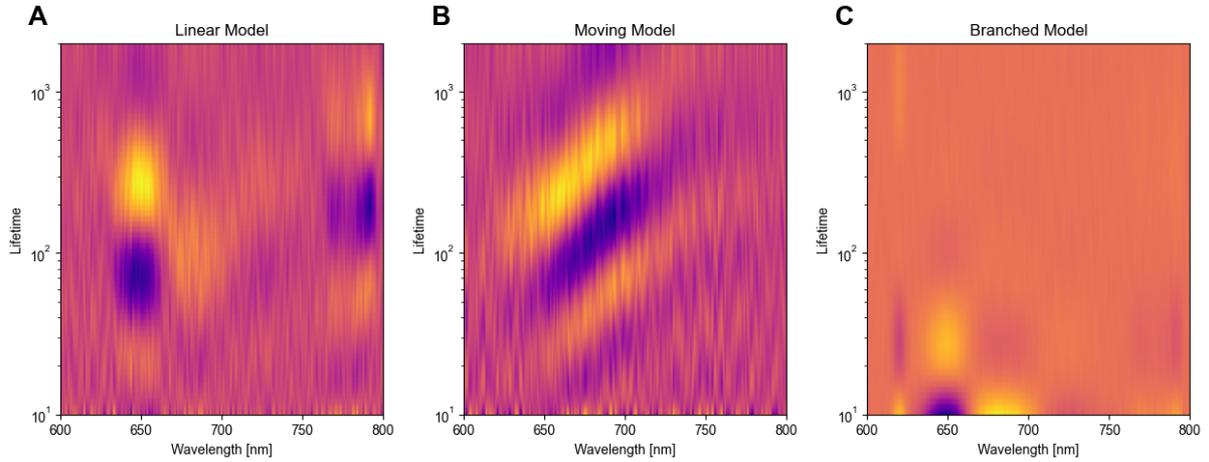

**Methods Figure 2: Lifetime Density Analysis of Synthetic Datasets**
Lifetime density maps obtained from analysis of **A -** sequential decay model, **B -** moving decay model, **C-** branching decay model. Negative values (blue) represent signal decay of a state at that lifetime whereas positive values (orange) represent signal growth of a state at that lifetime.

LDA in our implementation uses a large logarithmically distributed set of lifetimes, as proposed by Dorlhiac *et al.*[59], demonstrated in **Methods Figure 2** on the synthetic datasets. LDA attempts to find the vector of coefficients, $\vec{a}$, which minimizes,

$$||E(\vec{\tau}).\vec{a} - D||$$

*Methods Equation 11*

for design matrix $E(\vec{\tau})$. Here $D$ represents the transient signal at each measurement wavelength. However, as the number of lifetimes in LDA is large, it means over-fitting of the data is a bigger concern and so we utilise regularization. To compensate for over-fitting, we utilise regularization, whereby we amend the objective function by the addition of a penalty term designed to prevent extreme coefficients of fitting. With regularization, we minimize the amended function

$$\left\lVert \mathrm{E}(\vec{\tau}).\vec{a} - \mathrm{D} \right\rVert + f(\vec{a}),$$

where $f(\vec{a})$ is some regularization function giving rise to a penalty term to hinder spurious inputs of $\vec{a}$.

A number of regularization procedures exist, all imposing different assumptions on the data. Dorlhiac *et al.* give a detailed summary of the applicability of different regression algorithms in lifetime density analysis[59]. Two popular regularization algorithms are least absolute shrinkage and selection operator (LASSO) and Tikhonov both of which are implemented in our software.

Tikhonov regularization is useful when there is little known about the underlying kinetic distributions[59]. Tikhonov will produce results with somewhat broader kinetic distributions about key lifetime contributions. It is therefore difficult to separate two near-lying kinetic distributions, or a single lifetime from a nearby distribution. LASSO regularization can set coefficients to zero, whereas Tikhonov cannot. This makes the differences between the exponential coefficients for associated lifetimes greater. Thus, LASSO should be used where the data is expected to be described by few, narrowly distributed lifetimes [59].

Although there exist some metrics to help guide the choice in regularization coefficient, we fit to a wide number and allow the user to visually inspect the fit. As a default the algorithm will cycle through a hundred different regression coefficients and allow the user to pick the most effective.

## 11. Compartment Analysis

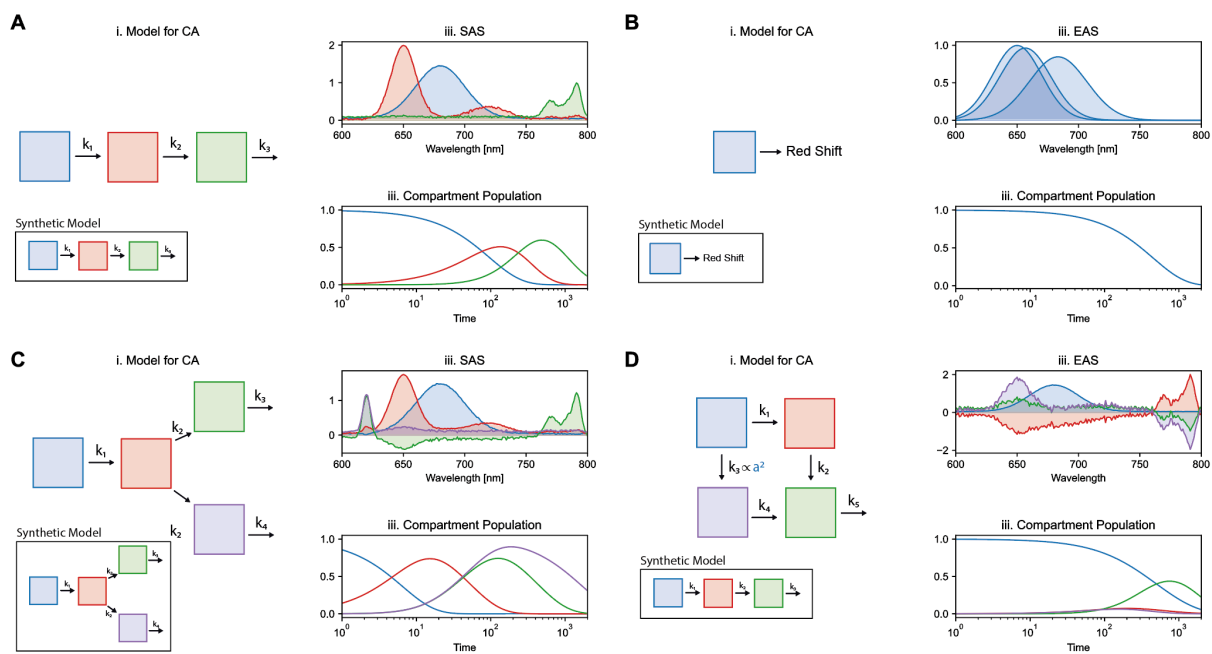

**Methods Figure 3: Compartment Analysis of Synthetic Datasets.** *Sub-panels depict (i) the kinetic models used, and (ii) species-associated or evolution-associated spectra, (iii) compartment concentrations over time obtained from CA. **D** highlights the difficulties in choosing a model, as we may also find good fit for the sequential decay models that are not representative of the original reaction scheme.*

In our implementation, the user may write an arbitrary differential model, which is then numerically solved, compared here between the synthetic models and chosen models in **Methods Figure 3**. The user may observe the residue in real time, and the optimisation may be parallelised across multiple computational cores if appropriate. The species-associated spectra[47] is determined by

$$\mathrm{SAS} = \mathrm{D}.\mathrm{C}^+,$$

*Methods Equation 13*

where $\mathrm{C}^+$ is the pseudoinverse of the concentrations of each component, recovered from the numerical model. Similarly to GA, the residue between the measured data and the model $\mathrm{SAS}.\mathrm{C}$ is minimised with respect to the lifetime.

# Acknowledgements

We acknowledge C. Faessler (Department of Plant Sciences, University of Cambridge) for providing a culture of *P. tricornutum*. Furthermore, we want to thank A. W. Rutherford (Imperial College London) for the gift of isolated PSII. T.K.B. gives thanks to the Centre for Doctoral Training in New and Sustainable Photovoltaics (grant no. EP/L01551X/2) and the NanoDTC (grant no. EP/ L015978/1) for financial support. D.K. thanks the Gates Cambridge Trust for financial support with support from the Benn W Levy Trust. J.D.C acknowledges the Waste Environmental Education Research Trust for funding. J.M.L. acknowledges financial support from the Biotechnology and Biological Sciences Research Council (BB/M011194/1) and the Worshipful Company of Leathersellers. L.T.W. acknowledges financial support from the Cambridge Trust. We acknowledge financial support the EPSRC (grant no. EP/M006360/1) and the Winton Program for the Physics of Sustainability.

# Open Access Statement

This work was funded by the UKRI. For the purpose of open access, the author has applied a Creative Commons Attribution (CC BY) licence to any Author Accepted Manuscript version arising.

# Author Contributions

T.K.B. initially developed the application of ultrafast techniques to examine cyanobacteria in collaboration with L.T.W. L.T.W, D.K. and J.M.L. cultured, prepared, and characterised the biological samples. J.C. cultured the *R. palustris*. H.M. purified the isolated PSI. B.W. isolated and cultured the *Sphingopyxis* isolate. C.J.H. supervised the biological work. T.K.B. performed the TA experiments with support from D.K. and J.M.L. T.K.B. developed the BioTADA software, performed the TA analysis, prepared the figures and wrote the initial draft with input from D.K. and J.M.L. All authors contributed to discussions, writing and editing of the manuscript.

# Data Availability Statement

The data underlying all figures in this article are publicly available from the University of Cambridge repository at (DOI to be added in proofs).